\documentclass[fleqn,10pt, twocolumn, preprint]{wlscirep}
\usepackage[utf8]{inputenc}
\usepackage[T1]{fontenc}

\usepackage{cleveref}
\usepackage{siunitx}
\usepackage{mathtools}
\usepackage[caption=false]{subfig}
\usepackage{enumitem}
\usepackage{multirow}
\usepackage{scalerel,amssymb}
\usepackage{verbatim}
\newcommand{\dbar}{d\hspace*{-0.08em}\bar{}\hspace*{0.1em}}

\UseRawInputEncoding 
\crefname{figure}{figure}{figure} 
\Crefname{figure}{Figure}{Figure} 
\crefname{table}{table}{table}
\Crefname{table}{Table}{Table}
\crefname{equation}{equation}{equation}
\Crefname{equation}{Equation}{Equation}

\usepackage{xcolor}

\title{Multi-scale model predicting friction of crystalline materials}

\author[1,2,+,*]{Paola C. Torche}
\author[1,2,x,*]{Andrea Silva}
\author[1,4]{Denis Kramer}
\author[1]{Tomas Polcar}
\author[1]{Ondrej Hovorka}
\affil[1]{Engineering and Physical Sciences, University of Southampton, Southampton, SO17 1BJ, United Kingdom}
\affil[2]{national Centre for Advanced Tribology Study, University of Southampton, Southampton, SO17 1BJ, United Kingdom}
\affil[4]{Mechanical Engineering, Helmut Schmidt University, Hamburg, 22043, Germany}
\affil[+]{pc.torche@soton.ac.uk}
\affil[x]{a.silva@soton.ac.uk}
\affil[*]{These authors contributed equally to this work}

\begin{abstract}
We present a multi-scale computational framework suitable for designing solid lubricant interfaces fully in silico. The approach is based on stochastic thermodynamics founded on the classical thermally activated two-dimensional Prandtl-Tomlinson model, linked with First Principles methods to accurately capture the properties of real materials. It allows investigating the energy dissipation due to friction in materials as it arises directly from their electronic structure, and naturally accessing the time-scale range of a typical friction force microscopy. This opens new possibilities for designing a broad class of material surfaces with atomically tailored properties. We apply the multi-scale framework to a class of two-dimensional layered materials and reveal a delicate interplay between the topology of the energy landscape and dissipation that known static approaches based solely on the energy barriers fail to capture.
\end{abstract}
\begin{document}

\flushbottom
\maketitle
%
%



\noindent
Recent years have witnessed an expansion of tribological research into "green tribology", which strives for the reduction of friction and elimination of environmentally toxic lubricants~\cite{nosonovsky_green_2010}. Achieving green tribology will require a systematic `bottom-up' design of new material surfaces with radically different physical properties. Among the promising candidates to achieve this goal is the class of materials with reduced dimensionality, such as two-dimensional layered transition metal dichalcogenides (TMD). Materials like WTe$_2$ and MoS$_2$ promise superior structural and mechanical properties that will lead to reduced dimensions, costs and increased efficiency in applications \cite{Manzeli2017}. The nanoscale friction properties of TMDs are thus actively studied~\cite{Vazirisereshk2020a}.

Innovative integration of the TMDs as coatings into a tribological design and assessing their energetic efficiency requires understanding their irreversible thermodynamic behaviour during the nanoscale friction processes and its extrapolation to macroscopic scales. This calls for a data-driven approach implementing efficient and accurate multi-scale modelling techniques to inform and interpret laboratory experiments.
These include the lateral-force atomic force microscopy (AFM), which is the key experimental tool used for quantifying the nanoscale friction processes ~\cite{Gnecco2001}. In particular, AFM is able to record the mechanical force exerted by a crystalline surface onto a nano-scale asperity dragged on top of it.
%
Common computational models frequently used to study the microscopic laws of friction include quantum-mechanical First Principles calculations, atomistic models based on molecular dynamics, nonlinear Prandtl-Tomlinson (PT) or Frenkel-Kontrova models, agent-based earthquake models, and models based on continuum mechanics applicable at macroscopic scales. The multi-scale modelling approach requires interfacing typically two or more of these levels of modelling into a systematic framework~\cite{Vakis2018}.

So far, the most refined multi-scale modelling of atomic friction has combined First Principles calculations with molecular dynamics methods \cite{Irving2017a, cammarata_atomic-scale_2019}.
%
While highly accurate, this methodology requires significant computational resources, which restricts its applicability to relatively short time- and length-scales. More importantly, the lack of reliable force fields for the majority of materials is a major limitation for the transferability needed for new-materials screenings.


The mesoscopic to macroscopic scale range of the friction processes has been studied by bridging atomistic molecular dynamics, linear response theory, and continuum mechanics into a unified multi-scale approach ~\cite{manini2016frictionRev}. However, continuum theories inherently exclude the possibility of thermal and structural fluctuations and their applicability to the nanoscale friction range becomes problematic, as it is inherently a far-from-equilibrium phenomenon dominated by such fluctuations and size effects  \cite{Luan2005}.

Instead, it is often more fruitful to employ non-equilibrium statistical mechanics combined with transition state theory or stochastic Langevin dynamics~\cite{zwanzig2001nonequilibrium}. This approach has been successful in generalising, for example, the classical PT model to describe the thermally activated nano-scale friction in AFM experiments~\cite{Prandtl1928, Tomlinson1929, Riedo2004, Krylov2005, Fajardo2014}, which qualitatively captured the velocity, load, and temperature dependencies observed in experiments~\cite{mate1987, Riedo2004, jansen2010}. To advance this statistical level of modelling requires incorporating the fundamental ability to describe the nanoscale frictional behaviour of specific materials, which is the main objective in this work.

The main result of this work is a fully consistent thermally activated thermodynamic model, which combines mesoscopic dynamics of a typical AFM tip sliding on two-dimensional (2D) material surfaces quantified through \textit{ab initio} calculations. The mesoscopic dynamics of an AFM tip was described by the adaptation of classical thermally activated PT model of dry adhesive friction~\cite{Fajardo2014} coupled to the framework of modern stochastic thermodynamics~\cite{crooks1999, Seifert2012}. This allowed us to systematically relate the randomised trajectories of the AFM tip to the corresponding fluctuating friction force, thermodynamic work, internal free energy, and entropy production $-$ an unambiguous measure of energy dissipation and microscopic irreversibility.
To obtain the potential energy surfaces to describe thermal actication of realistic materials, we performed systematic electronic structure calculations based on Density Functional Theory (DFT).
We characterised a class of two-dimensional materials, including graphene, hexagonal Boron Nitride (h-BN) and  TMDs monolayers, and also non-lubricating materials including two slabs of NaCl and a P bilayer in the As polymorph (P As-type), and succeeded in classifying these materials based on their nanoscale frictional performance characteristics.

We demonstrated that the developed multi-scale framework indeed allows evaluating the thermally activated frictional behaviour of different crystals directly from their respective electronic structure, which fulfils the key requirement for the bottom-up design of material surfaces with atomically tailored properties. Moreover, we illustrated that the combined multi-scale approach is essential for describing the fundamental aspects of the directional dependence of nanoscale friction not only in TMD material surfaces but also generally.

\section*{Results}
\begin{figure*}[!ht]
    \centering
    \includegraphics[width=0.9\textwidth]{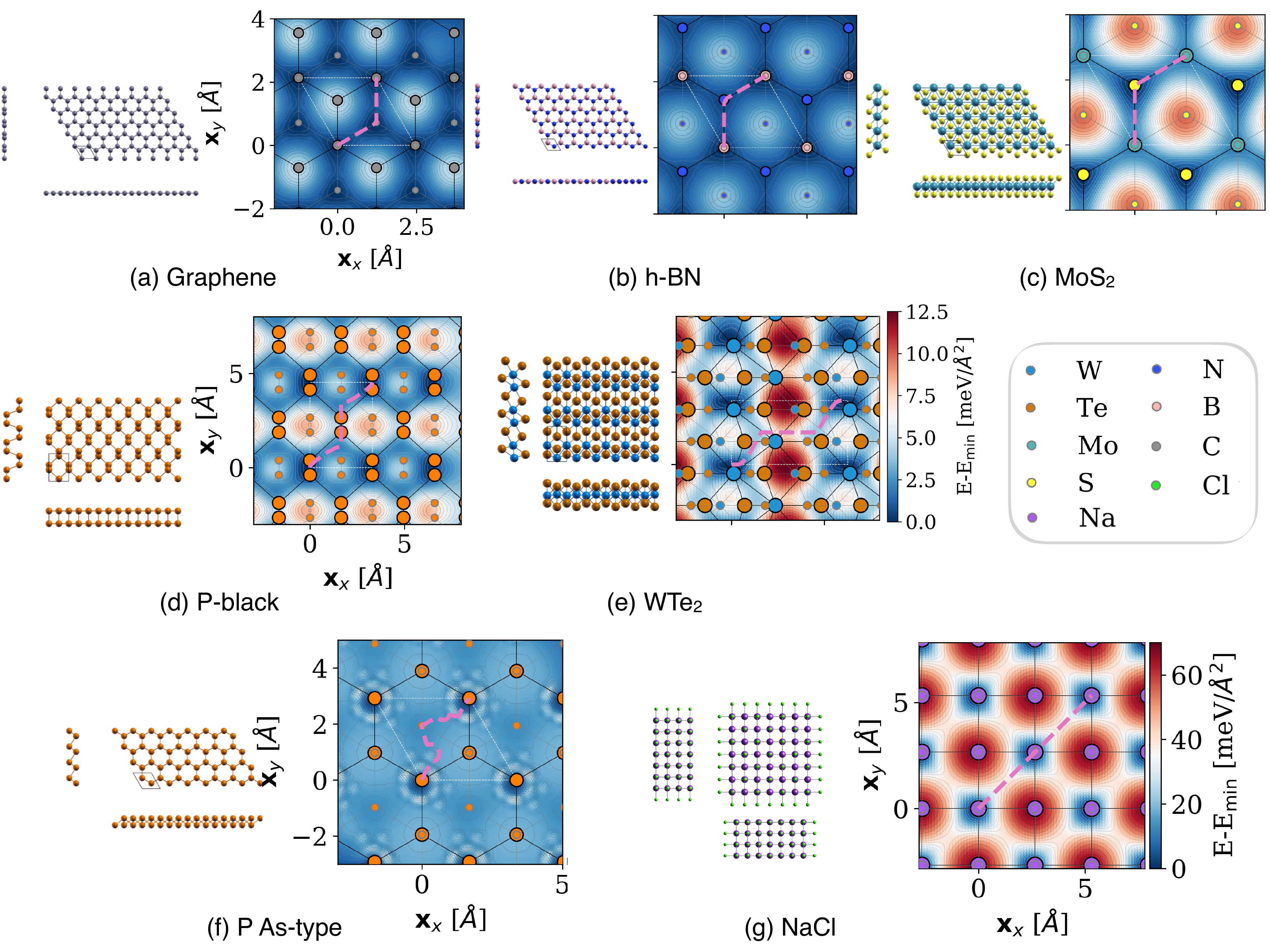}
    \caption{
    PES for the selected lubricant systems evaluated over a 15x15 grid at DFT level and interpolated over 200x200 points (see Supplementary Information Section II).
    Balls-and-sticks images on the left of each plot report the crystal structure of each ML system.
    The PES for the BL system is reported on the right.
    Smaller, gray-edged circles represent atoms in the bottom layer, while larger, black-edged circles represent atoms in the top layer.
    %
    %
    %
    The color bar reported in figure (e) refers to all figures (a)-(e). %
    Figures (f) and (g) show equivalent PES calculations for the selected non-lubricant systems. The corresponding energy scale is shown in the color bar in figure (g).
    Pink dashed lines in each PES plot indicate the MEP.
    The energy along each MEP is reported in Section III of the SI.
    }
    \label{fig:PES}
\end{figure*}
\bigskip

\noindent \textbf{PES for selected crystalline interfaces.} We studied a collection of 2D materials shown in \cref{fig:PES} with varying crystallographic structure and chemical composition relevant for solid lubrication technologies and green tribology, to ultimately assess the impact of complexity of their interface on energy dissipation.
To quantify the interaction between these materials and the AFM tip we used conventional approach combining layered DFT calculations with the Hertz model of the tip contact area \cite{Bhushan2005,Ouyang2020}. Specifically,
to obtain the potential energy surfaces (PES) $\mathcal{P}(\boldsymbol r)$ for these materials, decribing the energy corrugation per unit area, we used DFT calculations, as explain in detail in the Methods section and Section II of the SI.
%
The atomic scale corrugation as a function of the position in the material unit cell $\mathcal{P}(\boldsymbol r)$ was modelled at the DFT level as two flat crystalline surfaces sliding adiabatically, obtained from a set of translated infinite slab geometries.
The effect of the finite-size AFM tip was then included by renormalising the PES obtained in this way, in units of energy per area (meV/$\si{\angstrom}^2$), by the contact area between the tip and substrate via the Hertz model, yielding the static energy potential $U_\mathrm{s}$ in \cref{eq:energy} (in meV) as explain in the methods section.
%
This renormalisation is a crucial step in the construction of the multi-scale model as it allows mimicking the experimental protocol such as, a spherical tip coated with a thin crystalline layer in contact with a thin layer of the same crystalline material deposited over the substrate~\cite{Gao2015b, Liu2018}.

\Cref{fig:PES}(a)-(b) show PES calculations for purely 2D materials, namely two sliding graphene monolayers (a) and h-BN (hexagonal Boron Nitride) monolayers (b), which are both known for their excellent lubricant properties \cite{Vanossi2020a}. Both systems are composed of flat sheets of atoms arranged in a honeycomb lattice. \Cref{fig:PES}(c),(e) show PES for compounds whose monolayer geometry extends in 3D, which include two binaries from the TMD family, WTe$_2$ and MoS$_2$. These TMDs are composed of a transition metal layer sandwiched between two calchogenide planes, organised in prismatic coordination for MoS$_2$ (\cref{fig:PES}(c)) and in distorted octahedral coordination for WTe$_2$ (see \cref{fig:PES}(e)), which inherently gives them 3D structure. \Cref{fig:PES}(e) corresponds to the single-component black phosphorous P-black, which was recently identified as a promising solid lubricant~\cite{Losi2020}.
P-black consists of staggered rows of three-fold coordinated P atoms, with two in-plane neighbours and one in the neighbouring plane.
Finally, in \cref{fig:PES}(f)-(g) we considered non-lubricant materials P As-type, which is an hypothetical two-dimensional polymorph of P\cite{Mounet2018}, and a bilayer crystal composed of two slabs of NaCl, which has been used in AFM experiments to acquire high-contrast frictional maps~\cite{Socoliuc2006}.

The topology of the PES varies with compound coordination and chemistry.
The simplest energy landscape is found for graphene bilayer (GBL) in  \cref{fig:PES}(a), with a single global maximum and a single global minimum separated by a saddle point. The two equivalent minima in the unit cell corresponds to AB stacking of the carbon layers. They are shown as deep blue regions at the corners of the cell (shown in dashed white lines) and at the position $2/3 \mathbf{a}_1 + 1/3 \mathbf{a}_2$, with $\mathbf{a}_{1,2}$ being the unit vectors of the unit cell. The maxima, highlighted by the white region at $1/3 \mathbf{a}_1 + 2/3 \mathbf{a}_2$, corresponds to AA stacking.
%
%
\Cref{fig:PES}(b) shows a similar plot for h-BN, which has the same crystal structure as the graphene but it is composed of two different chemical species, B and N. This chemical heterogeneity within the same geometry leads to a more complex PES. The interactions between the B and N atoms lift the degeneracy of the two equivalent minima seen in the case of graphene bilayer, resulting in the appearance of a local minimum highlighted by the slightly lighter blue colouring at $1/3 \mathbf{a}_1 + 2/3 \mathbf{a}_2$. The global minimum corresponds to the AA stacking where B atoms sit on top of N atoms, while the local minimum is the AA stacking with the same atom types located on top of each other. The global maximum corresponds to the AB stacking, highlighted in \cref{fig:PES}(b) by white regions.

MoS$_2$, in \cref{fig:PES}(c), shows a PES with the same topology as h-BN, albeit with a higher corrugation, defined as the amplitude of the PES, ($\approx\SI{10}{meV}$) in comparison to h-BN ($\approx\SI{3}{meV})$.
In \cref{fig:PES}(d-e), the systems WTe$_2$ and P-black are characterised by an orthorombic cell and show the most complex PES, with local minima and maxima accompanying the global ones.

Finally, the energy landscape for the non-lubricant systems, NaCl and P As-type bilayers in \cref{fig:PES}(f) and (g), is one order of magnitude higher than those for the layered materials in \cref{fig:PES}(a)-(e), as highlighted in the associated colorbar, which supports the observations of higher friction in experiments~\cite{Socoliuc2006}.
For NaCl shown in \cref{fig:PES}(g), the high corrugation of about $\SI{60}{meV/\si{\angstrom}^2}$ arises from the strong electrostatic interactions between positively charged Na and negatively charged Cl ions. The energy minima correspond to Na-Cl atoms facing each other between the planes, shown by deep blue colouring, while the energy maxima correspond to same ions facing each others and are shown by deep red.
In the case of P As-type, the puckered structure of the layer, shown in \cref{fig:PES}(f), realises an atomic-scale egg-box geometry where in the stable configuration the two layers interlock each other, which corresponds to the deep-blue minima. A considerable energy is needed to break this lock, giving rise to the white ring of maxima around each minima seen in \cref{fig:PES}(f).

We note that the PES respects the symmetry of the unit cell in all cases, giving rise to triangular lattices of maxima and minima for hexagonal crystals, shown in \cref{fig:PES}(a)-(b), and (d), and rectangular lattices for orthorombic cells, shown in \cref{fig:PES}(c) and (e).

\begin{figure}[!ht]
\centering
    \label{fig:2DPT-example_pes3D}
    \includegraphics[width=0.9\columnwidth]{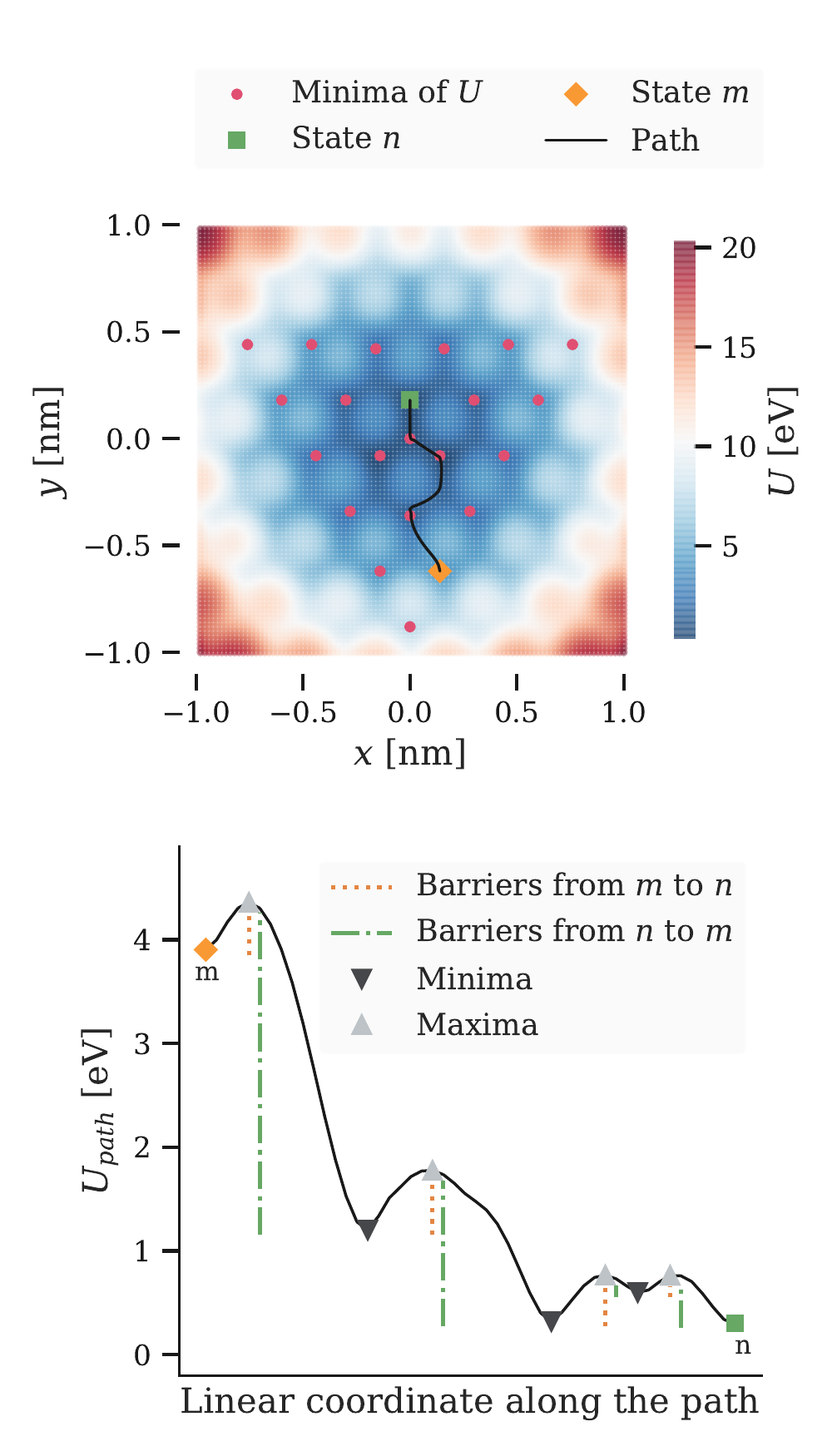}
    \caption{
    Example of total energy landscape $U$ and barrier between two states $m$ (orange circle) and $n$ (green circle) in MoS$_2$ at time $t_0=0$.
    The least energy path for the transition is found using the string method.
    We define partial energy barriers along the path as shown by the orange dotted, and green dot-dashed lines. See the methods section for details.
    }
    \label{fig:energy barrier}
\end{figure}

%
\bigskip
\noindent \textbf{Stochastic nanoscale friction model}. The potential energy of the system composed of the AFM tip sliding along a material surface can be written as:
\begin{equation}
\label{eq:energy}
    U(\boldsymbol r,t) = U_\mathrm{s}(\boldsymbol r)+\frac{C}{2} (\boldsymbol r-\boldsymbol v t)^2,
\end{equation}
where $C$ is the effective lateral compliance of the AFM contact, comprising the sample, tip and cantilever, and the vectors $\boldsymbol r$ and $\boldsymbol v$ are, respectively, the position of the tip and the sliding velocity of the cantilever in the $xy$-plane. The static energy surface, $U_\mathrm{s}(\boldsymbol r)$, describes the interaction of the tip with the material surface.
The surface $U_\mathrm{s}(\boldsymbol r)$ is obtained with the following renormalisation procedure, explained in details in the Methods section.
The atomistic PES $\mathcal{P}(\boldsymbol r)$ shown in \cref{fig:PES} (in energy per unit area), is multiplied by the contact area $A(L)$ obtained from the Hertz model at a given load $L$, yielding an energy per AFM tip.
The PES is considered independent of the load $L$, as the   typical loads accessible within AFM experiments cannot alter significantly the distance between MLs in contact ~\cite{Jacobs2013}.
Hence, the static energy landscape used as input to the PT model is $U_\mathrm{s}(\boldsymbol r,L) = \mathcal{P}(\boldsymbol r) A(L)$.

An example of the energy landscape defined by \cref{eq:energy} for an ideal energy surface  $U_s$ with hexagonal crystal symmetry is illustrated in \cref{fig:energy barrier}(a) (see also Supplementary Information Section I).
The total energy $U$ is shown by the coloured surface, with its local minima marked by circles.
The minimum energy path (MEP) between the states $m$ and $n$, which are associated with the energy minima corresponding to different locations at the surface of the sample, is highlighted and equals the path the tip is statistically most likely to take when transitioning from $m$ to $n$. Here, the MEP is identified by the string method~\cite{E2007}, due to its fast convergence. The actual energy profile along the highlighted MEP is shown in \cref{fig:energy barrier}(b), including the energy barriers, $\Delta U_{mn}$, which were identified by the tracking algorithm (see the methods section).
These energy barriers determine the transition rates between the states according to the Arrhenius law:
\begin{equation}
\label{eq:transition-rates}
\omega_{mn}=f_0 \exp \Big(-\beta \Delta U_{mn} \Big)
\end{equation}
where $f_0$ is the attempt frequency setting the characteristic timescale of thermal relaxation processes, $\beta=1/k_BT$, $T$ is the temperature, and $k_B$ the Boltzmann constant.
The thermally fluctuating sliding dynamics of the tip consistent with \cref{eq:energy,eq:transition-rates} is described by the master-equation introduced in the methods section, which has been solved here by standard time-quantified Monte-Carlo techniques to obtain the randomised long-timescale trajectories resembling those of an AFM tip in typical friction force microscopy experiments ~\cite{Riedo2004, jansen2010}.

Once the fluctuating trajectories have been computed, we have evaluated the corresponding stochastic thermodynamics. Stochastic thermodynamics extends the standard notions of thermodynamic work, entropy production, heat flow, etc., to small fluctuating systems~\cite{Seifert2012,VanDenBroeck2015}. It has been applied to microscopic systems such as elastic molecular systems~\cite{Shank2010}, nanomotors~\cite{Martinez2016}, colloidal particles in non-harmonic potentials~\cite{Blickle2006}, or to AFM in vertical harmonic oscillator mode~\cite{Gomez-Solano2010}, for example. Here we have applied the framework of stochastic thermodynamics to nanoscale friction by extending earlier work~\cite{pellegrini2019} to evaluate, besides the fluctuating work, $\Delta w$, also the irreversible heat, $\Delta q$, produced during the sliding motion of the AFM tip (see the methods section). This in turn allowed obtaining a detailed irreversible thermodynamic description of realistic materials, upon first establishing their PES and the corresponding substrate potential $U_\mathrm{S}$ in \cref{eq:energy} from First Principles calculations, as discussed above.

Finally, the tip-substrate potential $U_\mathrm{s}(\mathbf{r})$ entering in \cref{eq:energy} is obtained by renormalising the PES by the Hertz contact area (see Methods section). This then allowed obtaining the associated energy barriers (\cref{fig:energy barrier}), compute the thermally activated motion of the tip, and evaluate the stochastic thermodynamics over the resulting fluctuating tip trajectories, as discussed above.

\bigskip

\begin{figure}[!ht]
\centering
    \includegraphics[width=0.45\textwidth]{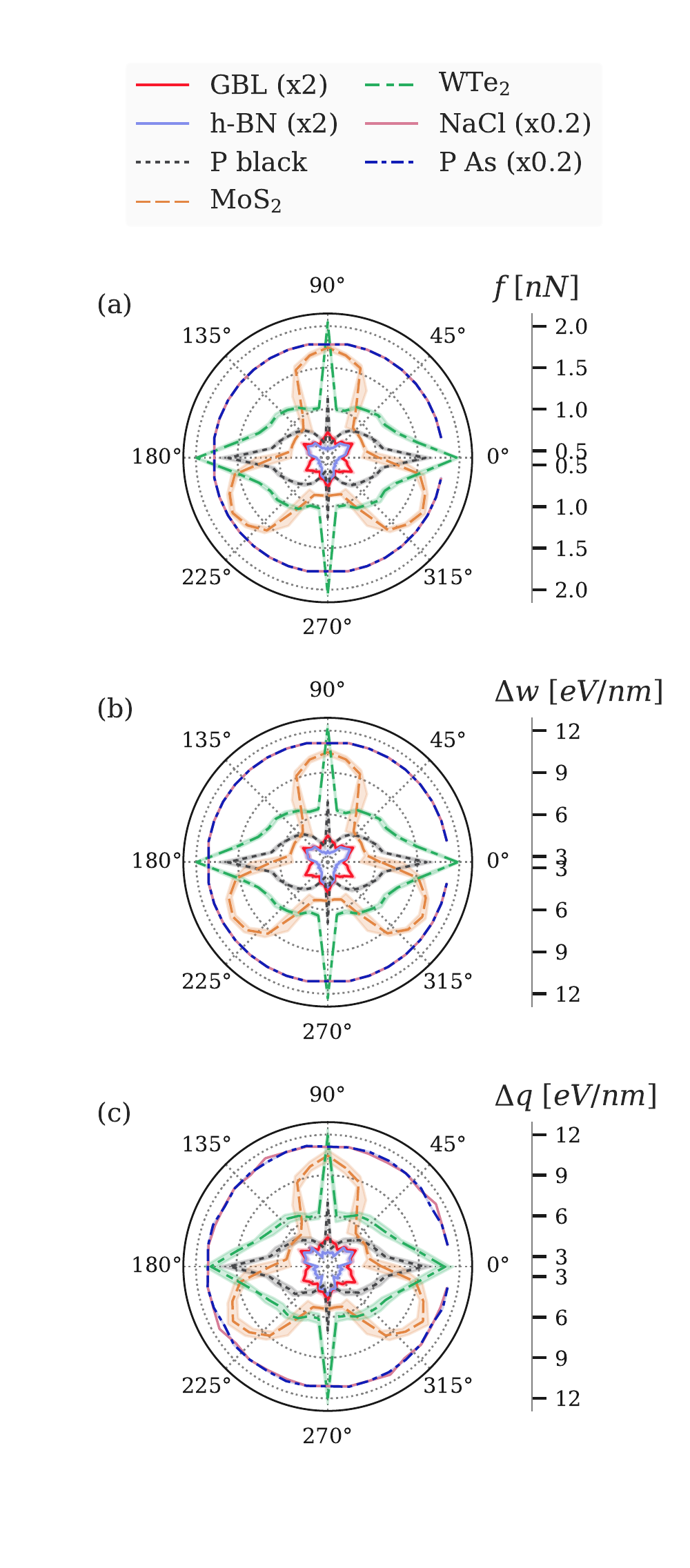}
    \caption{
    Angular dependence of the thermodynamic behaviour for different materials.
    The quantities shown are the time average of the lateral force after the first stick-slip event ($f$), the work of the lateral force ($\Delta w$), heat flow into the heat bath ($\Delta q$).
    All quantities are integrated over a sliding length of $\SI{20}{nm}$, and then normalised in units of eV/nm.
    All quantities are averaged over 50 trajectories.
    }
    \label{fig: anisotropy}
\end{figure}

The polar plots in \cref{fig: anisotropy}(a)-(c) show the computed friction force $f$, work $\Delta w$, and heat $\Delta q$ transferred to the heat bath for the different materials as a function of the sliding direction of the cantilever, in steps of $\ang{10}$ between the lattice direction and the cantilever velocity vector $\boldsymbol v$. The data shown is the average over 50 random trajectories for each material and sliding direction.
These calculations reveal the directional dependence of frictional behaviour with certain high and low friction directions highlighted, respectively, by large and small radial values in the plot. The high friction directions coincide with the increased energy (heat) dissipation and also the increased work expended during the dragging of the tip across the substrate. Note that the observed directional dependence of friction is especially pronounced for the TMD material class, which may act as good lubricants in certain directions and as non-lubricants in other directions.
Moreover, these plots allow to quantify to what extent the frictional force, the only quantity measurable in experiments, relates to the other thermodynamics quantities. In the case shown in \cref{fig: anisotropy}, we see that the angular dependence is the same for all quantities, meaning that the force is a good qualitative descriptor of the overall thermodynamics of the system.

The calculations for non-lubricants NaCl and P As-type do not display directional dependence. The simulations with the lubricant materials reach the steady-state of stick-slip motion in reasonable computational times, where the tip alternates long periods trapped around the potential minima and sudden jumps between these. Conversely, simulations with the non-lubricants show that the tip remains in the initial minimum instead of undergoing stick-slip motion. The reason is the large energy barriers for these materials (\cref{fig:PES}) relative to the elasticity constant $C$ of the cantilever in \cref{eq:energy}. Consequently, the estimates of thermodynamic variables in \cref{fig: anisotropy} were based on the initial energy barrier rather than the averaged stick-slip trajectory of the tip, and thereby relate to a lower bound of static friction. Increasing the parameter $C$ is necessary to observe the stick-slip motion for these two materials (see the methods section and Supplementary Figure 9).

\section*{\label{sec: discussion}Discussion}
\setlength{\arrayrulewidth}{0.2mm}
\setlength{\tabcolsep}{3.5pt}
\renewcommand{\arraystretch}{1.2}
\begin{table}
\small
\centering
\begin{tabular}{l|c|c|c}
         Material & $\langle f \rangle$  [nN] & $f_\mathrm{M}$ [nN] & $\nabla T$ in Si $[10^{-11}\mathrm{K/nm}]$   \\
            \hline
            \hline
            h-BN & 0.29$\pm$0.03 & 1.6 & 2.2 $\pm$ 0.3\\
            \hline
            GBL & 0.33$\pm$0.03 & 1.5 & 2.5 $\pm$ 0.2\\
            \hline
            P black & 0.91$\pm$0.24 & 1.5 & 6.7 $\pm$ 1.8 \\
            \hline
            WTe$_2$ & 1.30$\pm$0.31 & 3.4 & 9.4 $\pm$ 2.1 \\
            \hline
            MoS$_2$ & 1.67$\pm$0.43 & 3.7 & 9.0 $\pm$ 2.5\\
            \hline
            NaCl & \textit{9.00}$\pm$\textit{0.00} & \textit{15.6} & \textit{69.2} $\pm$ \textit{0.7} \\
            \hline
            P As-type & \textit{9.00} $\pm$\textit{0.00} & \textit{16.9} & \textit{69.2} $\pm$ \textit{0.4} \\
\end{tabular}
\caption{Materials ranked by average friction over all angles and temperature gradient in Silicon with thermal conductivity of $\lambda =\SI{8.11e11}{eV \cdot nm /(s \cdot K)}$.
The quantity $f_\mathrm{M}$ is the maximum lateral force obtained from the maximum shear strength in the material specific energy surface.
For the last two materials the simulations did not reach steady state due to large energy barriers and the values reported correspond to estimates based on the initial energy barrier. See the text for details.}
\label{tab:rank}
\end{table}

The selected crystals in \cref{fig:PES} can be ranked in terms of their lubricant character.
Integrating the profiles in \cref{fig: anisotropy} over all angles, we obtain the average value of the lateral force $\langle f\rangle$ reported in \cref{tab:rank}. h-BN and GBL are found to be the bests solid lubricants from the studied group, followed by  P-black, WTe$_2$, and MoS$_2$.
The non-lubricants NaCl and P As-type are identified as materials with poor lubricating properties, which corroborates earlier theoretical expectations and experimental observations~\cite{Socoliuc2006}.
Again, the estimates of the mean force $\langle f\rangle$ for these non-lubricant materials in \cref{tab:rank} are based on the initial energy barrier and relate to a lower bound of the static friction, due to the large energy barriers (see \cref{fig:PES}(f),(g)).

A partial validation of these results can be found in literature for MoS$_2$ and Graphene, the most studied materials of the selection.
The average friction force of $\SI{1.67\pm 0.43}{nN}$ for MoS2 predicted here is in line with the value of $\approx\SI{2}{nN}$ measured by Ky and coworkers \cite{Ky2018}.
The predicted friction for graphene $\SI{0.33 \pm 0.03}{nN}$ agrees with the value $\SI{0.306\pm 0.04}{nN}$ reported by Dienwiebel and coworkers \cite{dienwiebel2004}.
Both experiments are carried out at load ($\SI{10}{nN}$ in Ref.~\cite{Ky2018} and $\SI{18}{nN}$ in Ref.~\cite{dienwiebel2004}) comparable with the one adopted here ($\SI{10}{nN})$.

We compare the angle-averaged thermal friction force $\langle f\rangle$ to the static frictional force $f_\mathrm{M}$ (also given in \cref{tab:rank}), computed from the ideal shear strength determined directly from DFT. This represent the state-of-the-art friction descriptor at this level of description. The ideal shear strength is the largest negative value of the force obtained as a gradient along the minimum energy path and normalised by the Hertz contact area~\cite{Levita2015}. Thus, $f_\mathrm{M}$ is the estimate of the friction force based solely on the quantities obtained from First Principles calculations. It represents an upper bound for friction, as it is computed at 0 K and from static calculations. Due to the absence of thermal fluctuations, we expect $f_\mathrm{M} > \langle f\rangle$. Thus the present stochastic thermodynamic framework allows to evaluate the friction force subject to thermal fluctuations,  differentiate between the different sliding directions as in \cref{fig: anisotropy}, and to quantify heat production, which is an unambiguous measure of irreversibility. These are not possible with the estimations based on the shear strength variable, as far as the authors are aware.

A relevant application of the presented thermodynamic framework for friction is the estimation of contact temperature, which is of high importance for monitoring structural changes, but elusive and difficult to measure in experiments of Friction Force Microscopy. Here, we compute the temperature gradient produced during the sliding motion by integrating over the profiles of heat in \cref{fig: anisotropy}(a). The quantity $\Delta q$ in \cref{fig: anisotropy}(c) corresponds to the heat flow between the tip contact area and the heat bath, which can be used to estimate the local temperature gradient in the neighbourhood of the tip trajectory.
For instance, if the heat is transferred to the silicon core of the tip and the under-layer silicon substrate, then we obtain the temperature gradient as $\nabla T = v\Delta q /\lambda$, where $v$ is the magnitude of the velocity of the slider, and $\lambda$ the thermal conductivity of Silicon. The obtained values of the gradient for the investigated material selection, shown in \cref{tab:rank}, allows to estimate the heat generated in a frictional contact, a valuable information for engineering application.
\begin{figure}[t]
\centering
    \includegraphics[width=0.9\columnwidth]{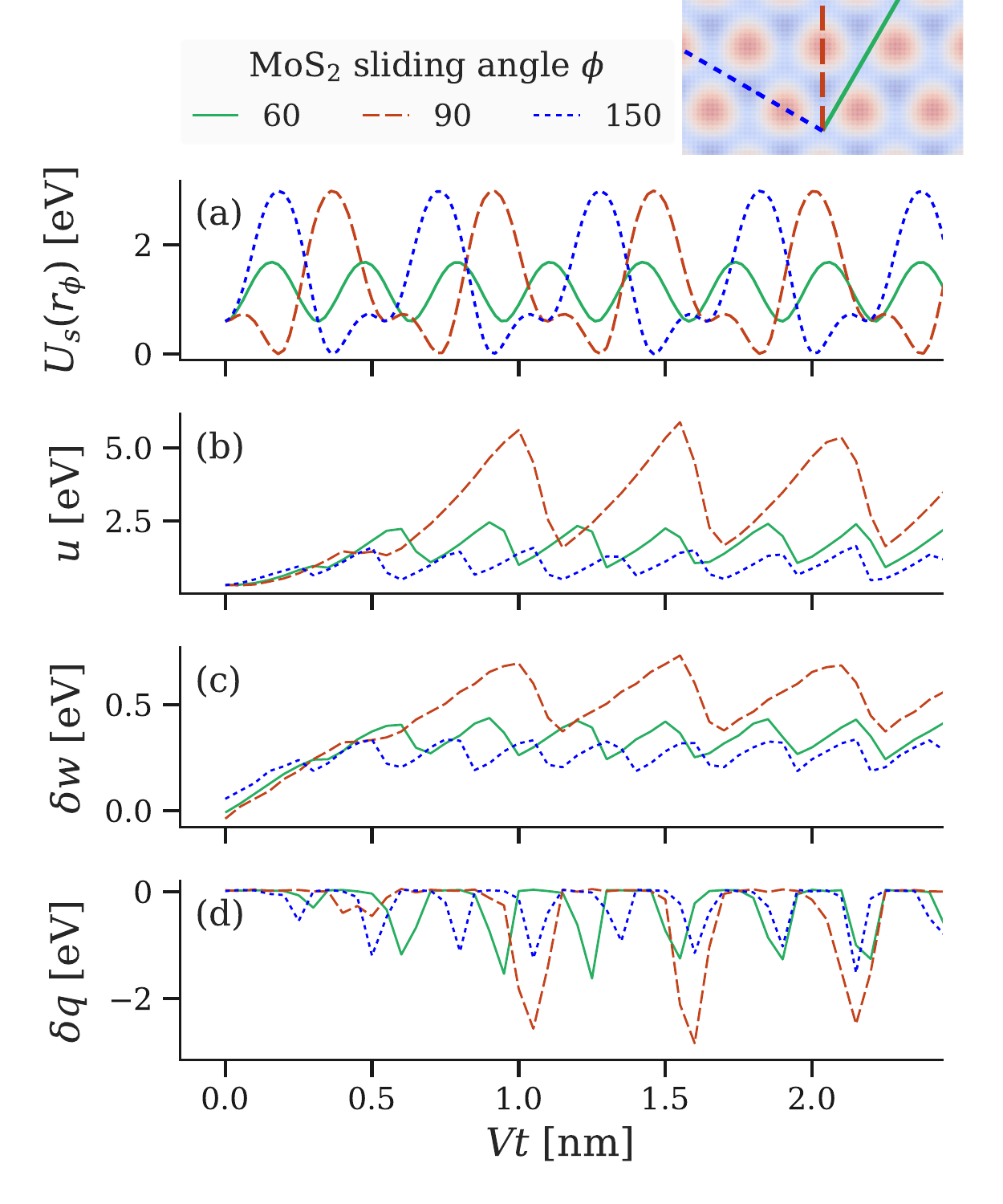}
    \caption{Instantaneous thermodynamics (a) substrate energy along the sliding direction, (b) fluctuating internal energy, (c) fluctuating work, and (d) fluctuating heat exchange with the heat bath, for MoS$_2$ sliding at $\ang{60}$ (green continuous), $\ang{90}$ (orange dashed) and  $\ang{150}$ (blue dotted). All quantities are averaged over 50 trajectories.
    The deviation from the mean is too small to be appreciable and has not been reported in the plot.
    }
    \label{fig: thermo_example_angles}
\end{figure}

The topology of the PES gives rise to the strongly directional nature of the thermodynamic behaviour observed in \cref{fig: anisotropy}.
To understand the relation between the PES geometry and lubricant properties, \cref{fig: thermo_example_angles} compares the stochastic thermodynamic behaviour in MoS$_2$ for three different sliding directions, namely $\ang{60}$,  $\ang{90}$, and  $\ang{150}$.
The static substrate energy $U_\mathrm{s}(\boldsymbol r)$ projected onto the sliding direction of the cantilever is shown in \cref{fig: thermo_example_angles}(a). However, the actual motion of the tip itself is not constrained to this direction and is allowed to fluctuate around it to follow the minimum energy path (see Supplementary Figure 7).
The resulting fluctuating internal energy is reported in \cref{fig: thermo_example_angles}(b).
\Cref{fig: thermo_example_angles}(c),(d) show the fluctuating work, and the heat transferred away from the system into the heat bath, respectively.

The figure shows a clear difference between the directions.
The trajectory along $\ang{60}$ (solid green line), cutting along the saddle points, yields the smallest dissipation.
On the other hand, the trajectories along $\ang{90}$ (dashed red line) and $\ang{150}$ (dotted blue line) are characterised by a similar 1D static landscape that interestingly results in a qualitatively different dissipation.
The large energy barriers in the substrate along the $\ang{90}$ direction as shown by the dashed line in \cref{fig: thermo_example_angles}(a) ultimately result in large internal energy changes produced during the sliding motion, significant work required to drive the system, and large dissipated heat bursts, as shown in \cref{fig: thermo_example_angles}(b)-(d).
Despite the similar substrate energy $U_\mathrm{s}$ along the $\ang{90}$ and $\ang{150}$ directions (\cref{fig: thermo_example_angles}(a)), the work and heat dissipation along the $\ang{150}$ direction is about half of that along the $\ang{90}$ direction (\cref{fig: thermo_example_angles}(b)-(d)). This behaviour can be attributed to the reflection symmetry of these energy barrier profiles, which  results in reversed ordering of the energy barriers visited by the tip during the motion due to the uni-directional driving.
%

%
\begin{figure}[t]
\centering
    \includegraphics[width=0.9\columnwidth]{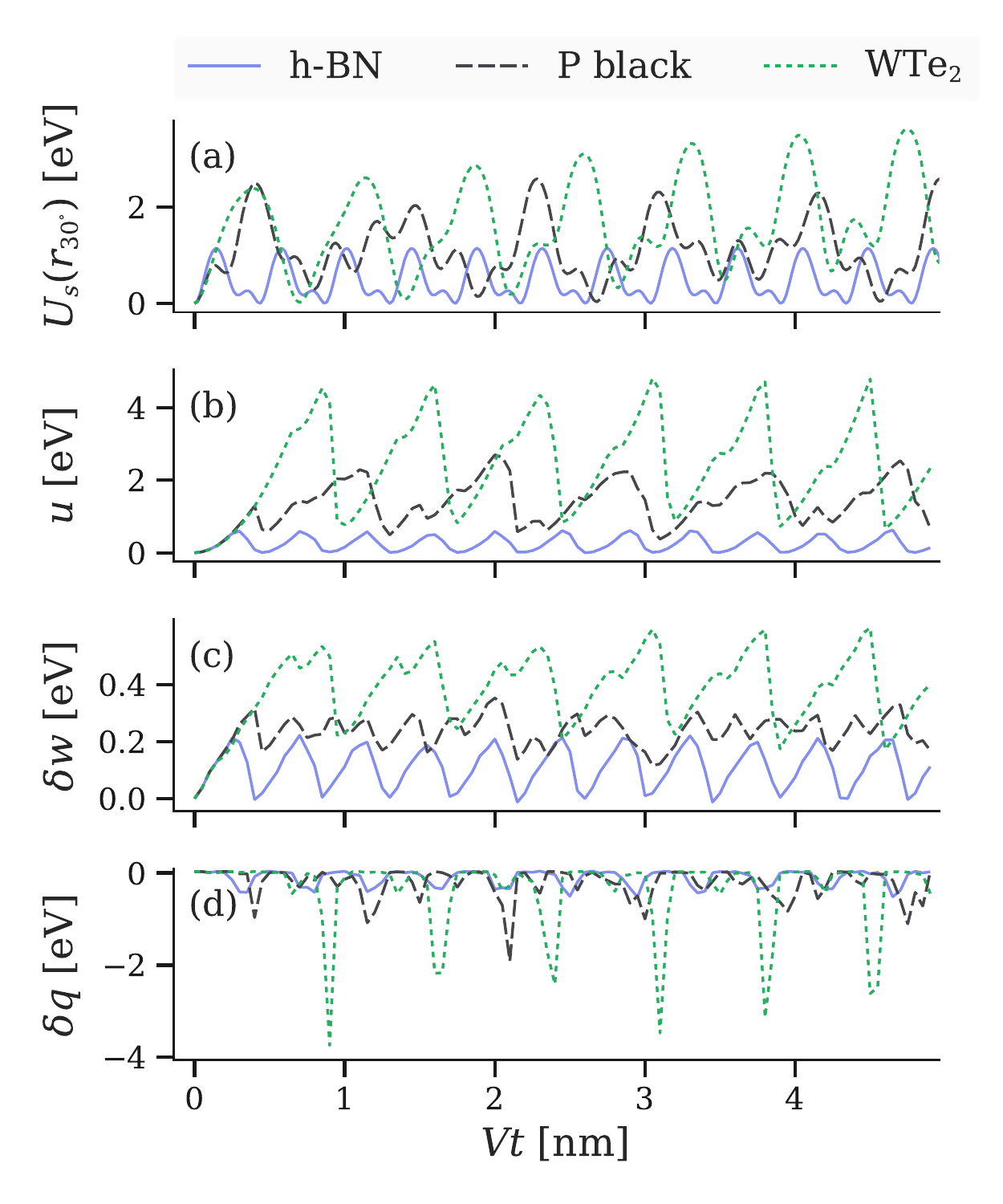}
    \caption{
    Instantaneous thermodynamics (a) substrate energy along the sliding direction, (b) fluctuating internal energy, (c) fluctuating work, (d) fluctuating heat exchanged with the heat bath for three different substrates, h-BN (blue continuous), P black (gray dashed), and WTe$_2$ (green dotted) sliding at $\ang{30}$.
    All quantities are averaged over 50 trajectories.
    The deviation from the mean is too small to be appreciable and has not been reported in the plot.
    }
\label{fig: thermo_example_substrates}
\end{figure}

A key achievement of our framework is to allow for systematic rationalisation of frictional behaviour of different materials.
As an illustrative example, in \cref{fig: thermo_example_substrates} we compare the behaviour of h-BN, P black, and WTe$_2$ for a fixed sliding angle, which we arbitrarily set to $\ang{30}$. The heat dissipation is the highest for the TMD material WTe$_2$ and the lowest for h-BN, as suggested by the size of the peaks in the work and the dissipated heat in \cref{fig: thermo_example_substrates}(c)-(d). An important result is that the static energy from the PES calculation (\cref{fig: thermo_example_substrates}(a)) alone is not necessarily an accurate descriptor of the extent of dissipative behaviour in TMD materials. The height and slope of the curves of P-black and WTe$_2$ in~\cref{fig: thermo_example_substrates}(a) are of similar value, but the work and the heat dissipation for WTe$_2$ shown in ~\cref{fig: thermo_example_substrates}(d) is nearly double that of P-black. Thus, employing the calculations based on the thermally fluctuating dynamics combined with the stochastic thermodynamics is strictly necessary for describing the frictional behaviour of these materials correctly.

The presented multi-scale approach eliminates the need for the computationally expensive molecular dynamics based modelling that is limited by a lack of force fields for the majority of materials. Two important applications can be highlighted. First, evaluation of friction and heat production for a material specific electronic structure, including fluctuations and anisotropy effects. And second, prediction of temperature increase in the contact. This last is usually eluded in experimental studies of nanoscale friction and not possible to measure in Friction Force Microscopy. However, it is of major importance, as it may trigger structural changes and chemical reactions in contact surfaces~\cite{Nicolini2018a}.

Experimental validation of the present results could be achieved, for example, by using the recently developed experimental protocol~\cite{Liu2018}, which allows wrapping flakes around AFM tips to yield the homostructured contact such as modelled here. By changing the relative orientation between the flakes and the tip before the wrapping, the mismatch angle between a sliding direction and crystal orientation can be varied and directional force maps such as the ones presented in \cref{fig: anisotropy} can be measured experimentally.
Finally, from the prospective of a bottom-up approach to solid lubricants design, our model can be integrated with existing material databases\cite{Jain2013a,Mounet2018} or PES databases\cite{Restuccia2018} to perform High-throughput screenings for material surfaces with enhanced lubricating properties. This systematic approach would allow to created a much needed frictional database for machine-learning aided materials discovery\cite{Zaidan2017} and for synthesis efforts.
%

\section*{Methods}

\noindent\textbf{Thermally activated model.} The minima of energy in \cref{eq:energy} define the notion of stable states $m$ of the system. The state energies are $U_m = U(\boldsymbol r_m, t)$, where $\boldsymbol r_m=\boldsymbol r_m(t)$ is the position of the tip in the state (energy minimum) $m$. Both, $U_m$ and $\boldsymbol r_m$, as well as the total number of states $m$ are time-dependent quantities, due to the relative motion of the tip. The rate of the transition from a state $n$ to a state $m$ is given by the Arrhenius law in \cref{eq:transition-rates}. The path between the states $n$ and $m$ is identified by the string method~\cite{E2007} in the two-dimensional total energy surface $U$ defined in \cref{eq:energy}. To identify the actual energy barriers $\Delta U_{mn}$ along this path, we first found local maxima and local minima along the path using a peak detection algorithm~\cite{2020SciPy-NMeth}, and defined $\Delta U_{mn}$ as a sum of partial energy barriers along the path in the direction from $n$ to $m$. This definition naturally implies higher energy barriers between distant states, and thus reduced likelihood of thermally activated transitions, in comparison to neighbouring states. The energy barriers $\Delta U_{mn}$ as well as the rates $\omega_{mn}$ are time-dependent.

Each state $m$ has an associated probability of occupation $p_m$, the evolution of which is given by the Master equation for Markovian dynamics:
\begin{equation}
\label{eq: Master equation}
    \frac{dp_m}{dt}=\sum_n \big( \omega_{mn}p_n - \omega_{nm}p_m \big)
\end{equation}
The stochastic trajectories, $n(t)$, defined as a set of states visited by the AFM tip during a given time interval, and an initial condition, can be found by Monte Carlo sampling of discrete states, as described in Supplementary Information Section IV.

To obtain the plots in \cref{fig: anisotropy,fig: thermo_example_angles,fig: thermo_example_substrates} we computed the randomised sliding tip trajectories using the Monte-Carlo method with parameters in \cref{eq:energy} in the range of values of a typical AFM experiment~\cite{dong_analytical_2011}: $C=\SI{3}{N/m}$, $|\boldsymbol v|=\SI{10}{nm/s}$,
$T=\SI{300}{K}$, and $f_0=\SI{10}{kHz}$.
The load was fixed at $L=\SI{10}{nN}$, yielding a contact area of $A=\SI{3.04}{nm^2}$, according to the Hertz model.
The sliding distance was set at $\SI{20}{nm}$, to ensure that the simulations reached the steady-state of stick-slip.

\bigskip


\noindent\textbf{Stochastic thermodynamics.} We applied the definitions of stochastic thermodynamics to the AFM tip trajectories $n(t)$ found from the Monte Carlo sampling method introduced in the model section.
The friction force over a stochastic tip trajectory $\boldsymbol r_{n(t)}$ is computed as $\boldsymbol f=-C(\boldsymbol r_{n(t)} - \boldsymbol v t)$, and then averaged over the length of the sliding interval~\cite{Gnecco2000}.
The fluctuating work of the cantilever per unit of time is then:
\begin{equation}
    \frac{\dbar w}{dt} = \boldsymbol f \cdot \boldsymbol v
\end{equation}
The path energy corresponding to the trajectory can also be evaluated from \cref{eq:energy} as $u=U(\boldsymbol r_{n(t)},t)$.
Assuming the system is embed in a constant temperature surrounding or a heat bath, from the first law of thermodynamics for microscopic closed systems~\cite{Sekimoto1998}, we can calculate the heat transfer to the system from the heat bath as:
\begin{equation}
\label{eq:first law}
\frac{\dbar q}{dt} =  \frac{du}{dt} - \frac{\dbar w}{dt}
\end{equation}
The notation $\dbar$ is used to emphasize variables dependent on the path, while $d$ is used for state variables. Additional thermodynamic variables such as fluctuating entropy, and entropy production are defined in Supplementary Table II~\cite{Seifert2012, torche_stochastic_2021}.

In simulations, we observe periodic behaviour of the energy and entropy of the system, after the initial transient period.
The choice of the starting and final position of sliding determines the cumulative change of energy and entropy which are zero if the system returns to the same state since they are thermodynamic state variables. The small change in internal energy observed in \cref{fig: anisotropy} is due to the start and end point of the tip during the sliding located at non-equivalent positions along the substrate.

\bigskip

%
\noindent\textbf{Density functional theory (DFT) calculations.} At the DFT level, the atomic scale corrugation is modelled as two flat crystalline surfaces sliding adiabatically.
Thus, the potential energy surface (PES) is obtained from a set of translated geometries as explained in detail in the Supplementary Information Section II.
Each translation point represents a DFT calculation of a bilayer system of the given crystal, e.g. MoS$_2$ on MoS$_2$, where the top layer has been translated by a vector $\bf{x}$ with respect to the bottom layer.
We assume the sliding interface comprises of several layers, a common situation in experiments \cite{Liu2018}.
To mimic the presence of more layers in the interface, the geometry is relaxed keeping the bottom layer fixed in $xyz$-directions and letting the top layer relax in the $z$-direction only, to avoid it sliding back to the minimum position.
Allowing relaxation along the $z$-direction corresponds to the sliding interface at zero load.

The calculations were carried out using the Vienna \textit{Ab initio} Simulation Package (VASP)~\cite{Kresse1993} within the Projector Augmented-Wave (PAW) framework~\cite{Blochl1994}.
The exchange-correlation potential was approximated using the PBE functional~\cite{Perdew1996} and the vdW dispersion was described by the DFT-TS method, a local-geometry-corrected empirical model developed by Tkatchenko and Scheffler~\cite{Tkatchenko2009}.
This dispersion has been shown to capture correctly the vdW interactions in layered materials in both the out- and in-plane directions \cite{Ouyang2020a,Marom2010}
A plane wave cut-off of $\SI{650}{eV}$ was adopted and the Brillouin zone was sampled using a $17\times17\times1$ mesh.

\bigskip


\noindent\textbf{Hertz model of the tip contact area.} The atomic-scale sliding interface and tip deformation determining the contact area are assumed to be decoupled processes.
The sliding interface arises from short-range interaction and, thus, is determined by properties of the 2D layers in contact~\cite{Jacobs2013}.
On the other hand, the contact area is determined by the deformation of the tip and the substrate and is usually modelled by continuum mechanics, which explicitly disregards the discrete nature of nanoscale contacts~\cite{Jacobs2013}.
An accurate description of the contact area in nanoscale tribology remains the subject of current debate~\cite{Mo2009,Gnecco2001}.
Here the Hertz model is used to estimate the contact area.
For fixed load, the contact area represents a scaling factor of the DFT-computed PES and is independent of the coating monolayer crystal.
Its value cannot change the relative amplitude and geometry of corrugation between different materials.
Thus, while the Hertz model may not describe the contact area exactly, with systematic deviation of up to 20\%~\cite{Jacobs2013,Luan2005}, it is suitable to understand qualitative trends and compare results within this coherent reference, as done in this work.
Not that while in this work the simplest solution is chosen to keep the proof-of-concept as clear as possible, the modular structure of the framework allows to exchange the Hertz model with DMT or JKR without changing any other part of the protocol.
Hence, more complex contact models can be adopted if the experimental setup requires it, while the Hertz model represents the clearest way to present the results.

In the Hertz model, the radius $a$ of the contact area is expressed in terms of the applied load $L$ and mechanical properties of the tip~\cite{Johnson2013} by:
\begin{equation}
\label{eq:Hertz_rad}
    a = \left(\frac {3LR}{4E^{*}}\right)^{1/3},
\end{equation}
where $R$ is the curvature radius of the tip and the effective elastic modulus is given by
\begin{equation}
    \frac {1}{E^{*}} = \frac {1-\nu _{1}^{2}}{E_{1}} + \frac {1-\nu _{2}^{2}}{E_{2}}.
\end{equation}
The quantities $E_1, E_2$ are the elastic moduli and $\nu_1, \nu_2$ the Poisson's ratios associated with the tip and material surface.
The quantities appearing in \cref{eq:Hertz_rad} are evaluated for silicon, a common material for tip and substrates.
Thus, $E_1=E_2=E$ and $\nu_1=\nu_2=\nu$.
The parameters $E=\SI{151.34}{GPa}$ and $\nu=0.19732 $ are obtained from DFT calculations, tabulated in the Materials Project~\cite{Jain2013a}.
%
The radius $R=\SI{10}{nm}$
of the AFM tip
is taken from experiments~\cite{Gnecco2001}.
Thus, the static potential $U_\mathrm{s}(\mathbf{r})$ in \cref{eq:energy} is obtained by scaling the infinite-interface PES computed using DFT as outlined above, scaled by the realistic tip-substrate contact area $A(L) = \pi a(L)^2$, where $a$ is the Hertz radius given by \cref{eq:Hertz_rad}.

\section*{Acknowledgements}
This project has received funding from the European Union's Horizon2020 research and innovation programme under grant agreement No. 721642: SOLUTION.
The authors acknowledge the use of the IRIDIS High Performance Computing Facility, and associated support services at the University of Southampton, in the completion of this work.

\section*{Author contributions statement}
P.T. and A.S. performed the simulations. D.K., T.P., and O.H. supervised the work. All authors reviewed the manuscript.

\end{document}